# Proactive Web Server Protocol for Complaint Assessment

G. Vijay Kumar[1]
M.Tech (CSE), (Ph.D)
Associate Professor
Department of CSE, ASTRA,
Hyderabad, INDIA

Ravikumar S. Raykundaliya[2]
(M.Tech (IT)),
Department of IT, ASTRA,
Hyderabad, INDIA

Dr. P. Naga Prasad[3]
M.Tech (CSE), Ph.D
Professor of CSE and
Dean Academics, ASTRA,
Hyderabad, INDIA

**Abstract-**-*Vulnerability Discovery with attack Injection security threats are increasing for the server software, when software is developed, the software tested for the functionality. Due to un-awareness of software vulnerabilities most of the software before pre-Release the software should be thoroughly tested for not only functionality reliability, but should be tested for the security flaws (or) vulnerabilities. The approaches such as fuzzers, Fault injection, vulnerabilities scanners, static vulnerabilities analyzers, Runtime prevention mechanisms and software Rejuvenation are identifying the un-patched software which is open for security threats address to solve the problem "security testing". These techniques are useful for generating attacks but cannot be extendable for the new land of attacks. The system called proactive vulnerability attack injection tool is suitable for adding new attacks injection vectors, methods to define new protocol states (or) Specification using the interface of tool includes Network server protocol specification using GUI, Attacks generator, Attack injector, monitoring module at the victim injector, monitoring module at the victim machine and the attacks injection report generation. This tool can address most of the vulnerabilities (or) security flows.*

## 1. INTRODUCTION

Network security consists of the provisions and policies adopted by a network administrator to prevent and monitor unauthorized access to the network, misuse, modification, or denial of a computer network and network-accessible resources. Network security involves the authentication of access to data in a network, which is managed by the network administrator.

Network is of two types i.e. private and public. Private network doesn't need Internet connectivity. So networks designed in this way are considered safe from Internet attacks, a part from those internal threats still exists with the development of huge open networks, Threats related to security have grown rapidly in the past few years. Hackers or Intruders have found out more network vulnerabilities, and because of the applications that can be downloaded freely and that require less or probably no hacking knowledge, troubleshooting and maintaining deliberated applications and optimizing networks if gone to the wrong hands can be used destructively and cause severe threats.

In computer security, vulnerability is considered as weakness or flaws that allow an attacker to lessen a system's information assure. Vulnerability is the union of three elements: a system flaw, flaw accessed by attacker, and attacker capability to handle the flaw. To find vulnerability, attackers should have at least one applicable tool or technique that can help to find system weakness. Here, vulnerability can be called as the attack surface.

A risk of security can be called as vulnerability. Uncertainty can be led by the application of vulnerability. The risk is considered as a potential of a significant loss. When the affected asset has no value is considered as vulnerabilities without danger. Vulnerability with number of known instances of working and fully implemented attacks is defined as an exploitable vulnerability, vulnerability for which can use the existing.

## 2. LITERATURE SURVEY

A literature review is a text written by someone to consider the critical points of current knowledge including substantive findings, as well as theoretical and methodological contributions to a particular topic.

2.1 Buffer Overruns

*Overview of the Vulnerability*

In low-level language buffer overrun is recognized as a problem. The main problem is that program flow control information and user data are mixed for the sake of performance, and low-level languages allow direct access to application memory. Buffer overrun affects the two popular languages c and c++.

Buffer overrun is affects crash to the attacker gaining complete control of the application and if a high-level user using the application (root, administrator, or local system), then control of users who are currently logged on to a system and the operating system being logged on, or will log on, will be handled by the attacker. Buffer overrun is first exploited by an internet worm to the finger server and is called as the Robert T. Morris (just Morris) finger worm. Even though it seems that we have learned how to ignore buffer overruns since the one which had nearly knocked down the Internet in 1988, continues reports of buffer overruns in different types of software are been referred.

*Spotting the Vulnerability Pattern*

Below are the components to look at:

- Input, whether it is read from the network, from the command line or from a file.
- Data being transferred from meant input to internal structures.
- Unsafe string handling calls are used.

*Testing Techniques to Find the Vulnerability*

Fuzz testing, which means intending your application to use semi-random inputs, is recognized one of the effective testing techniques to use. Increasing the length of input strings while observing the behavior of the applications are tried. The mismatches between input checking will result in relatively small windows of vulnerable code is what deserved sometime.

2.2 Format String Problems

*Overview of the Vulnerability*

Format string problems are one of the few truly new attacks to surface in recent years. One of the first mentions of format string bugs was on June 23, 2000.

As with many security problems, the root cause of format string bugs is trusting user-supplied input without validation. In C/C++, format string bugs can be used to write to arbitrary memory locations, and the most dangerous side is that this can happen without tampering with adjoining memory blocks. On Windows system, application string tables are generally kept within the program executable, or resource Dynamic Link Libraries (DLLs). If an attacker can rewrite the main executable or the resource DLLs, the attacker can perform many more straightforward attacks than format string bugs.

*Spotting the Vulnerability Pattern*

Any application that takes user input and passes it to a formatting function is potentially at risk. One very common instance of this vulnerability happens in conjunction with applications that log user input. Additionally, some functions may apply formatting internally.

*Testing Techniques to Find the Vulnerability*

Pass formatting specifies into the application and sees if hexadecimal values are given back. For example, if you have an application that expects a file name and returns an error message containing the input when the file cannot be found, then try out giving it file names like NotLikely%x%x.txt. If you get an error message along the lines of "NotLikely12fd234104587.txt cannot be found," then you have found format string vulnerability.

2.3 Integer Overflows

*Overview of the Vulnerability*

Integer overflows, underflows, and arithmetic overflows of all types, especially floating point errors, have been problem vulnerability the beginning of computer programming. Theo de Radar, of OpenBSD fame, claims integer overflows are "the next big threat." The writers of this book think we are at least three years into the threat.

*Spotting the Vulnerability Pattern*

Any application performing arithmetic can exhibit this vulnerability, especially when one or more of the inputs are provided by the user, and not thoroughly checked for validity. Focus especially on C/C++ array index calculations and buffer size allocations.

*Testing Techniques to Find the Vulnerability*

If the input is character strings, try feeding the application sizes that tend to cause errors. Like, strings that are 64K or 64K–1 bytes long can often cause problems. Other common problem lengths are 127, 128, and 255, as well as just on either side of 32K. Any time that adding one to a number results in either changing sign or flipping back to zero, you have a good test case.

**3. SYSTEM ANALYSIS**

Firewalls follow some simple rules such as to deny or allow protocols, IP address or port. Some attacks such as DoS attacks are too complex even for today's firewalls. The attack cannot be prevented by firewall because they don't have the capability to differentiate between good traffic from DoS attack traffic. Routers which are places in between may be affected even before the firewall gets the traffic.

Flooding is a kind of Denial of Service (DoS) attack that is designed in order to bring a network or service down by flooding it with huge amounts of traffic. Flood attacks take place when a network or service becomes too weighted with packets initiating incomplete connection requests that restricts genuine connection requests. The first packet sent across a TCP connection is known as a "SYN" or "synchronize" packet. For example, when you contact http://www.google.com, the first packet your systems out will be a SYN packet to the HTTP port 80 on www.google.com. Your browser is telling the web server that it wants to connect. A SYN flood is a form of denial-of-service attack in which an attacker sends a succession of SYN requests to a target's system in an attempt to consume enough server resources to make the system unresponsive to legitimate traffic. Data Tempering is to modify by the addition of a moderating element. It is necessary to understand that all methods of data transmission can be easily toggled by a user and that user data cannot be considered as reliable.

A penetration test is a method of evaluating computer and network security by simulating an attack on a network or computer system from external and internal threats. The process involves a vital analysis of the system, improper system configuration or poor result is produced from potential vulnerabilities that could result from the system, Software flaws or operational faults and known/unknown hardware in process of technical preventions. This observation is taken from the point of a potential attacker and can include activities exploitation of security vulnerabilities. Security issues resulting from the penetration test are represented to the system's owner. Effective penetration tests will collect this information with a proper assessment of the potential impacts to the organization and outline a range of procedural countermeasures and technical to reduce risks.

Penetration tests are valuable for several causes:
- Attack vector determines the Possibility of a particular set.
- Combination of lower-risk vulnerabilities exploited in a particular sequence results in finding high risk vulnerabilities.
- With the help of automated network or application vulnerability scanning software identifying vulnerabilities that may be difficult or impossible to detect with.
- Assessing the importance of potential business and operational impacts of successful attacks.

The Payment Card Industry Data Security Standard (PCI DSS), and security and auditing standard, requires both yearly and ongoing penetration testing are the examples of penetration tests which is a component of a full security survey.

## 4. SYSTEM DESIGN

In System design we are considering the step as we are going to exploit the Vulnerability using following parameters.

**Protocol Specification**
- The state-of-art tools use protocol specification as manual approach.
- In this work, proposing automatic extraction of protocol specification based on packet captures called pcaps.
- Pcaps can be gathered from live traffic flows from client to server.

**Attack generator**
- A TCP Session or a flow can be identified based on SIP,SP,DIP,DP,Protocol
    E.g.: 10.0.0.1, 1010, 10.0.0.2, 80, tcp.
- The module gathers individual flows for each session.
- Separates Request/Responses.
- Identifies buffer overflow producing patterns from the requests.
- Replaces patterns with buffer overflow payloads.

**Attack injector**
- The attack injector executes each test case
- Interacts with the monitor for the activity of the server status (crash, hang, etc,).
- The module executes all the test cases one-after the other.

**Target system and monitor**
- The module continuously monitors the status of the server.
- E.g.: If the server is crashed, it informs to the attack injector module.

**Vulnerability Report**
- The attack injector executes the attack script.
- The script sends the attack traffic towards the server.
- If the server hanged or crashed due to the attack traffic, the monitor confirms the status.
- If status is as expected, the report is prepared as bug report.

## 5. SYSTEM ARCHITECTURE

A System Architecture is the conceptual model that defines the structure, behavior, and more views of a system, organized in a way that supports reasoning about the structures of the system.

The state-of-art tools use protocol specification as manual approach. Proposing automatic extraction of protocol specification based on packet captures called pcaps. Pcaps can be gathered from live traffic flows from client to server. The module gathers individual flows for each session. Separates request/responses. Identifies buffer overflow causing patterns from the requests. Replaces patterns with buffer overflow payloads.

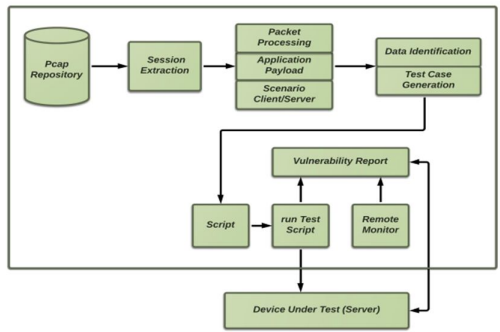

Fig 1: System Architecture

The module prepares executable attack scripts. Whereas each script as a test case. The process is repeated for all the flows. The attack injector executes each test case. The module executes all the test cases one-after the other. The module continuously monitors the status of the server. If the server is crashed, it informs to the attack injector module. The script sends the attack traffic towards the server. If the server hanged or crashed due to the attack traffic, the monitor confirms the status. If status is as expected, the report is prepared as bug report.

## 6. IMPLIMENTATION
**Proactive Approach**

We are first using Wire shark tool to capture packets and then we are Creating Database Tables to store the packet information, String type test cases and integer type test cases.

Protocol Specification, we are analyzing the Pcaps and find the packet number, source ip, source protocol, destination ip, destination protocol and insert this packet information into a table called packetflow.

Data Type Identification, we are differentiating string data type and integer data type from the captured packets. Here w indicates string and indicates integer value.

Test Case Generation, we are creating test cases for string data which we got into the packet. From the packet we will use packet no, start index and end index from the Pcap. And it will be replaced by string Payload.

Attack Generation, we are checking the ip address of client if the ip address is of client then it will attack and if it is of server means local host then it will skip.

We are using FTP Cesar and it a freely available on internet. FTP Server is having an easy to use interface; it allows you to set up an anonymous account in less than a minute. Although its interface is quite easy, it comes with advanced configuration options like firewall settings, user/group controls, statistics, bandwidth control, remote control of the server and others. Talking about security, Cesar FTP lets you ban users based on IP address or hostnames, all connection attempts can be limited then ban or kick user.

In General Cesar FTP is a quick, reliable solution for an FTP server need.

- Easy to use.
- Full control of users and what they are doing.
- Free.

Attack Generation, we are check that the ip address is of client or not if it of client it will create the payload for string and also for integer and it of server it will skip and find other ip addresses and send the payload data to the client ip.

Attack Injector, one array is created and it will store the injecting test cases and store into Status.dat file

Remote Monitor, test case is monitored by the remote monitor by using test case no and which test case is running and after 10 seconds it will be terminated or it will not be opened and if it will server is crashed then it will be killed by monitor.

## 7. CONCLUSION

We proposed a proactive web server protocol approach to detect vulnerability from web server so that confidential data. The penetration test case study that we have implemented effectively pivot through discovered ftp server parameters to achieve its goal. An effective security policy must limit at the lowest possible this information leakage. Additionally, system and service configurations must be carefully revised in order to implement only the necessary features, preventing critical information exposures. We have analyzed a penetration testing case study against a simulated network setup. Despite this up-to-date security policy, we have managed to compromise the internal network, taking advantage of mis-configurations and security design flaws. In existing system, it takes protocol specification as manual and through GUI. The testers are unaware of the protocol specifications causing usability issues. Proposed system solves the problem with pcaps. In this proposed work we have replaced Pcap data with the desired payload and able to crash the server by using buffer overflow concept to get vulnerability from server. This can be used to prevent the useful data from the organization. And in future this kind of proactive approach will be used by bigger companies to prevent their networks.

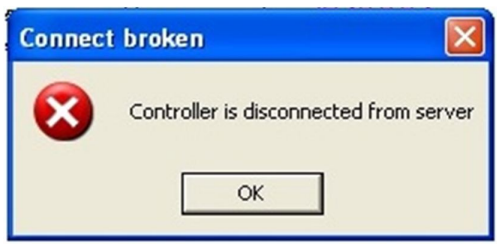

## REFERENCES

[1] McGraw, G. (2006). Software Security: Building Security In, Adison Wesley Professional.

[2] The Canadian Institute of Chartered Accountants Information Technology Advisory Committee, (2003) "Using an Ethical hacking Technique to Assess Information Security Risk", Toronto,Canada.http://www.cica.ca/research-and-guidance/documents/it-advisory committee/item12038.pdf, accessed on Nov. 23, 2011.

[3] Mohanty, D. "Demystifying Penetration TestingHackingSpirits,"http://www.infosecwriters.com/text_resources/pdf/pen_test2.pdf, accessed on Nov. 23, 2011.

[4] "Application Penetration Testing," https://www.trustwave.com/apppentest.php, accessed on Nov. 23, 2011.

[5] [Online]. Available. http://en.Wikipedia.org/

[6] [Online]. Available. http://www.google.co.in./

First Author: G Vijay Kumar, Associate Professor, CSE Dept, 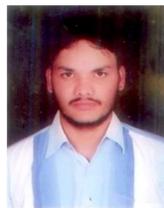 Aurora's Scientific Technological and Research Academy, completed his M. Tech (CSE) in 2008 and presently pursuing Ph.D (CSE) in Software Engineering from JNTU University, Hyderabad. He published several research papers in the field of Software Engineering and Computer Networks. His areas of research are including Software Engineering, Neural Network and Computer Networks.

Second Author: Ravikumar S Raykundaliya received his 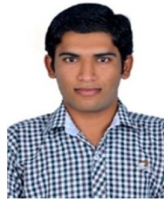 Bachelor of Engineering (BE) degree from Sant Gadge Baba Amravati University in 2011. He is currently pursuing M. Tech. in Information Technology, from ASTRA, Bandlaguda affiliated from Jawaharlal Nehru Technological University, Andhra Pradesh. His research interests are in Information and Network Security, Cloud Computing.

Third Author: Dr. P. Naga Prasad, Professor, CSE Dept, Dean 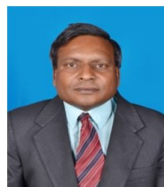 Academic, Aurora's Scientific Technological and Research Academy, has completed Ph.D in the area of Artificial Intelligence. He has published several research papers in Artificial Intelligence, Fuzzy Logic, and Software Engineering & Computer Network. He guided many students at M.Tech & Ph.D level; His research area includes Artificial Intelligence, Cloud Computing and Security, Software Engineering and Computer Network and Data Warehouse Systems.